\documentclass[a4wide]{PoS}

\usepackage{url}
\usepackage{caption}
\usepackage{subcaption}

\newcommand{\es}{1ES~0229+200}

\newcommand{\fermi}{Fermi--LAT}

\title{Unexpected $\gamma$--ray signal in the vicinity of \es}

\ShortTitle{Unexpected $\gamma$--ray signal in the vicinity of \es}

\author{Stanislav Stefanik and Dalibor Nosek \\
        Charles University, Faculty of Mathematics and Physics \\
        V Holesovickach 2, 180 00 Prague 8, Czech Republic\\
        E-mail: \email{stefanik@ipnp.troja.mff.cuni.cz}
        }

\abstract{
We report on an unidentified $\gamma$--ray signal found in the region around the BL Lac object~\es.
It was recognized serendipitously in our analysis of 6.2 years of \fermi~data at a distance less than $3^\circ$ away from the blazar.
The observed excess of counts manifests itself as an unexpected local maximum in the test statistic map.
Although several \fermi~sources have been identified in this area we were not able to link them to the position of this residual signal.
A clear association with sources visible in other wavebands was not successful either.
We briefly discuss characteristics of this unresolved phenomenon.
Our results suggest a steep energy spectrum and a point--like nature of this candidate $\gamma$--ray emitter.
}

\FullConference{The 34th International Cosmic Ray Conference,\\
		30 July -- 6 August, 2015\\
		The Hague, The Netherlands}

\begin{document}


\section{Introduction}

The third Fermi Large Area Telescope catalogue (3FGL) of high energy $\gamma$--ray sources contains 3033 sources found during 4 years of operation of the instrument~\cite{3FGL}.
Out of these, 1009 objects are unidentified as no counterparts at other wavelengths have been associated with them.
More sources are yet to be revealed since \fermi~has been conducting observations for almost 7 years.

In this work, we present our finding of a previously undocumented source of a signal in the \fermi~data.
It emerged during our analysis of a region around the BL Lac \es, our initial source of interest.
Moreover, several other fainter objects were revealed along that.
The analysis and its results are described in Section~\ref{sec:analysis}.
The discussion of the features of the found $\gamma$--ray excesses is given in Section~\ref{sec:discussion}.


\section{Analysis}
\label{sec:analysis}

We used the publicly available Pass 7 data from a region of interest (ROI) with a radius of 15$^{\circ}$.
The centre of ROI was at the position of \es~at $\mathrm{RA} = 02\mathrm{h}~32\mathrm{m}~48.6\mathrm{s}$, $\mathrm{Dec} = +20^{\circ}~17'~17.5''$.

We examined the data using the standard analysis chain of Fermi Science Tools (v9r33p0)~\footnote{\href{http://fermi.gsfc.nasa.gov/ssc/data/analysis/software/}{http://fermi.gsfc.nasa.gov/ssc/data/analysis/software/}}.
The chosen time range encompasses events registered between August 4, 2008 and October 31, 2014.
Events with energies between 100~MeV and 100~GeV were selected.
A cut on the maximum zenith angle of 100$^{\circ}$ was applied.
After the preliminary selections, we proceeded with the binned analysis and used the recommended instrument response functions  P7REP\_SOURCE\_V15.

The contribution from nearby sources to the observed emission in the ROI was accounted for by adding their spectral and spatial descriptions to the source model.
The diffuse background was modelled using the templates \texttt{gll\_iem\_v05\_rev1} and \texttt{iso\_source\_v05\_rev1} for the galactic and isotropic diffuse emission, respectively.
The model also contained known sources up to 25$^{\circ}$ from the centre of the ROI in order to take into account the point spread function (PSF) of \fermi~at low energies~\cite{PSF}.
The list of sources together with their spectral and spatial specifications was initially taken from the second \fermi~catalogue (2FGL) made after 2 years of data taking~\cite{2FGL}.
All spectral parameters were kept fixed for very faint sources and for those which lie beyond 7$^{\circ}$ from the centre of the ROI.
Since \es~is not present in the 2FGL catalogue, we added another source in the model at the centre of ROI and assumed its spectral shape is a simple power law.

Maximum likelihood estimates (MLE) of relevant parameters were found using the \emph{gtlike} tool.
In the preliminary analysis, we kept in the model only those sources, the contributions of which to the emission in the ROI were not negligible.
In the end, the blazar \es~was firmly recognized in the data with the test statistic $\mathrm{TS} = -2 \left( \mathrm{log}\mathcal{L}_0 - \mathrm{log}\mathcal{L} \right)=70$.
Here, $\mathcal{L}$ and $\mathcal{L}_0$ are the likelihood functions for the model with and without the additional source included, respectively.


\subsection{Source identification}
\label{sec:identification}

\begin{figure}[t]
	\begin{center}
		\includegraphics[width=0.8\columnwidth]{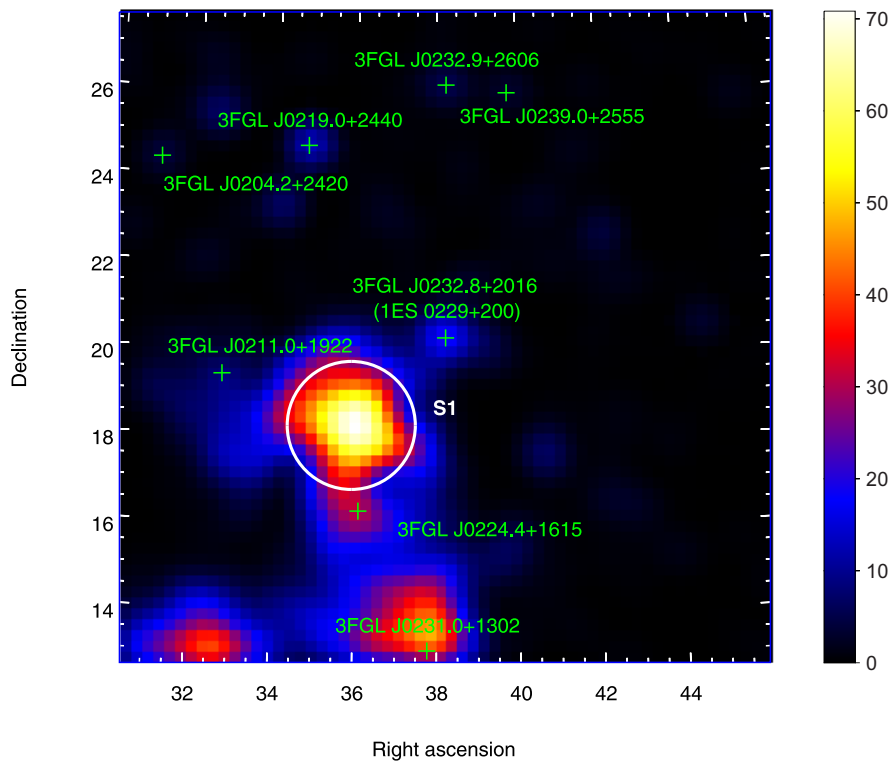}
		\caption{
			A map of test statistic values for bins of $0.25^{\circ}$ size in a $15^{\circ}\times15^{\circ}$ region centred around the position of \es.
			Only sources present in the 2FGL catalogue were assumed to contribute to the $\gamma$--ray signal in the shown region.
			Vertical bar at the right side represents colour encoding for different TS values.
			Gaussian smoothing was applied to the map.
			Labelled crosses denote sources present in the 3FGL but not in the 2FGL catalogue. 
			White circle marks the unexpected signal labelled as S1 unattributable to any catalogue source.
		}
		\label{fig:tsmap1}
	\end{center}
\end{figure}

\begin{figure}[t]
	\begin{center}
		\includegraphics[width=0.8\columnwidth]{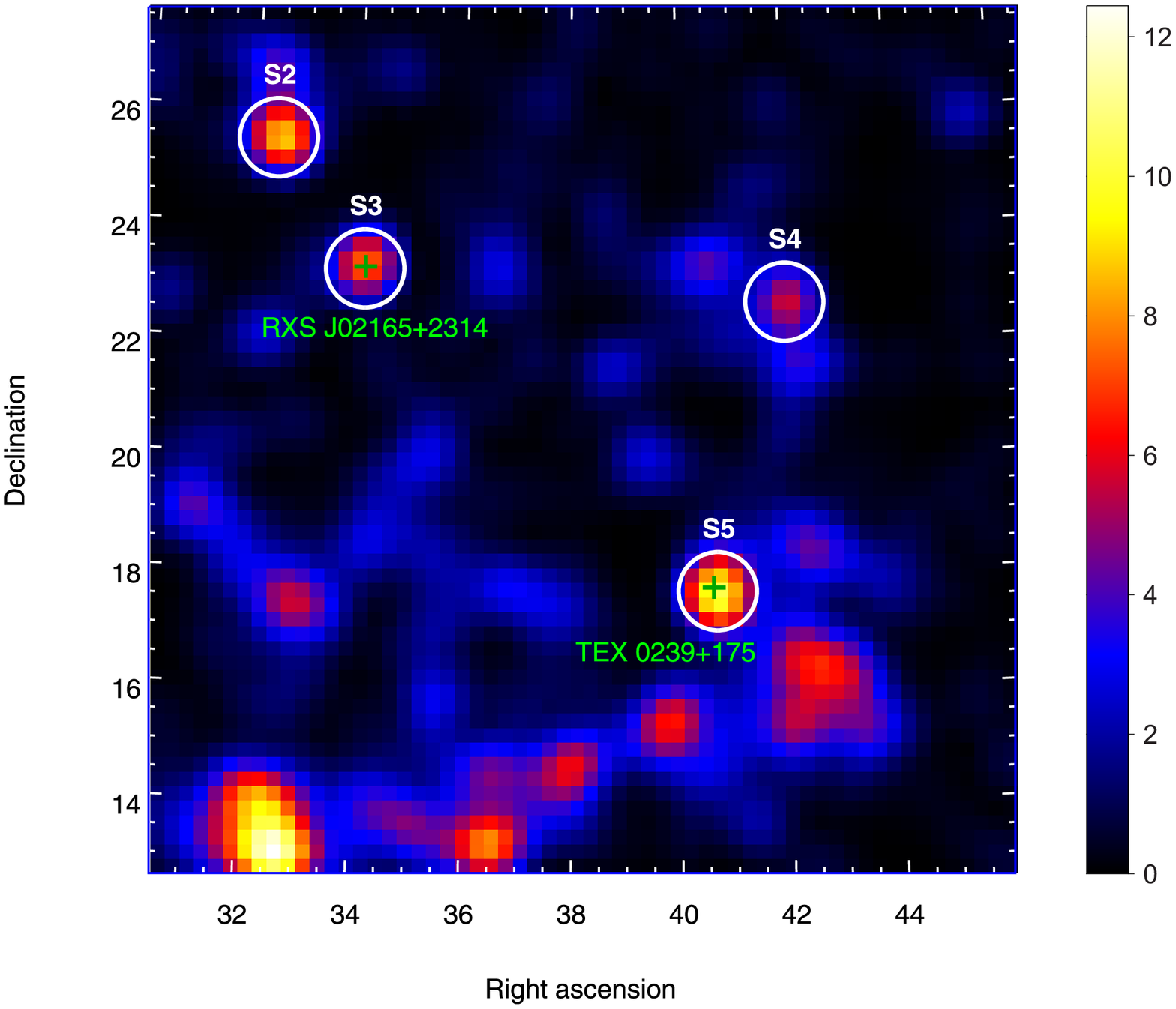}
		\caption{
			A map of test statistic values based on a source model including sources from the 3FGL catalogue and the hypothetical object S1.
			Circles encompass the unknown sources of HE $\gamma$--rays. 
			Green crosses pinpoint the positions of two known BL Lac objects~\cite{VCV}.
			For further details see caption to Fig.~\protect\ref{fig:tsmap1}.
		}
		\label{fig:tsmap2}
	\end{center}
\end{figure}

We checked for any additional sources which were not included in our source model but, nonetheless, contribute significantly to the observed signal in the investigated region.
For this purpose, we constructed a map of TS values based on the output model file which was obtained after the iterations of the \emph{gtlike} analysis.
\es~was excluded from this model.
We used the \emph{gttsmap} tool which calculates the TS values for individual spatial bins by introducing a putative source at a given position.
The map covers an area of $15^{\circ}\times15^{\circ}$ and its binning was chosen as $0.25^{\circ}$ per pixel.

The TS map is shown in Fig.~\ref{fig:tsmap1}.
The area in the lower left corner of the map is the only spot of residual TS values that can be linked to a source from the 2FGL catalogue.
We believe that this bright area is a relic of 2FGL~J0211.2+1050 which is located beyond the boundary of the map.
Although this source was included in our model, we presume that it appeared in the map due to its flaring activity in 2010 and 2011.

In Fig.~\ref{fig:tsmap1}, the presence of \es~in the data is visible as the spot in the centre of the map.
Additional bright areas appear in the TS map which do not correspond to any 2FGL counterpart.
After the release of the 3FGL catalogue~\cite{3FGL}, we compared the positions of these hotspots with the coordinates of sources reported in the 3FGL but not in the 2FGL catalogue (green crosses in Fig.~\ref{fig:tsmap1}).
In the investigated region, most of the emitters from the newer catalogue can be easily paired with the sources which appear in our TS map.
Note that the catalogue name of \es~is 3FGL~J0232.8+2016.

The area labelled as S1 in Fig.~\ref{fig:tsmap1} cannot be attributed to any of the known HE $\gamma$--ray sources.
This was confirmed after we rerun the analysis with the new source model made out of all 3FGL sources situated within the $25^{\circ}$ radius.
S1 remained the only clearly distinguishable residual signal.

In the next step, we included an additional source in our model.
Its spectral type was assumed to be a power law and the coordinates were read off from the TS map as the approximate position of the S1 hotspot.
After the MLE of the spectral parameters were found, \emph{gtfindsrc} tool was employed to find the best estimate of the position of the presumed source.
The resulting coordinates of S1 along with the TS value associated with this additional source are given in Table~\ref{tab:positions}.

In Fig.~\ref{fig:tsmap2}, a TS map obtained for the source model with S1 included is shown.
No other significant residuals can be recognized around its position.
However, four more spots became distinguishable at levels of significance sufficient for claims of detection (white circles in~Fig.~\ref{fig:tsmap2}).
We performed the same analysis routine as described above for the four new source candidates.
In Table~\ref{tab:positions}, the coordinates and TS values of these objects are listed.

Finally, the \emph{gtlike} tool was retried with a model containing all 3FGL sources as well as five objects S1--S5.
All of them were assumed to be point--like with power law shaped spectra.
The results of the analysis for the new objects are given in Table~\ref{tab:spectra}.
Spectral parameters of \es~are also stated therein.

\begin{table}
	\centering
	  \begin{tabular}{l c c c}
		\hline
		Object & TS & RA [$^{\circ}$] & Dec [$^{\circ}$] \\
		\hline
		S1 & 90.3 & 36.09 & 18.53 \\	
		S2 & 36.3 & 32.51 & 25.31 \\	
		S3 & 41.2 & 34.13 & 23.25 \\	
		S4 & 31.6 & 42.03 & 22.54 \\	
		S5 & 34.0 & 40.62 & 17.57 \\	
		\hline
	  \end{tabular}	  
	\caption{
	  	Test statistics associated with the detection of five unknown sources and their positions.
	  }
\label{tab:positions}
\end{table}

\begin{table}
	\centering
	  \begin{tabular}{l c c c}
		\hline
		Object & $N_0$ & $\Gamma$ & $I~(>100~\mathrm{MeV})$ \\
		& $(\mathrm{ph}~\mathrm{cm}^{-2}~\mathrm{s}^{-1}~\mathrm{MeV}^{-1})$ & & $(\mathrm{ph}~\mathrm{cm}^{-2}~\mathrm{s}^{-1})$ \\
		\hline
		S1   & $\left( 4.03 \pm 0.67 \right) \times 10^{-10}$    & $2.81 \pm 0.09$ & $\left( 2.22 \pm 0.30 \right) \times 10^{-8}$   \\	
		S2   & $\left( 1.71 \pm 1.06 \right) \times 10^{-11}$    & $1.87 \pm 0.16$ & $\left( 1.96 \pm 0.90 \right) \times 10^{-9}$   \\
		S3   & $\left( 1.40 \pm 0.89 \right) \times 10^{-11}$    & $1.81 \pm 0.16$ & $\left( 1.72 \pm 0.79 \right) \times 10^{-9}$   \\
		S4   & $\left( 1.62 \pm 1.15 \right) \times 10^{-11}$    & $1.86 \pm 0.17$ & $\left( 1.88 \pm 0.99 \right) \times 10^{-9}$   \\		
		S5   & $\left( 1.14 \pm 0.60 \right) \times 10^{-10}$    & $2.41 \pm 0.18$ & $\left( 8.09 \pm 3.36 \right) \times 10^{-9}$   \\
		\es   & $\left( 4.80 \pm 1.76 \right) \times 10^{-11}$   & $2.04 \pm 0.13$ & $\left( 4.61 \pm 1.31 \right) \times 10^{-9}$   \\
		\hline
	  \end{tabular}	
	  	\caption{
	  	Maximum likelihood estimates of spectral parameters of analysed objects.
	  	Differential energy spectra are given by the power law prescription as $F(E) = N_0 \left( E/E_0 \right)^{-\Gamma}$.
	  	Here, $N_0$ is the flux normalization at $E_0 = 100~\mathrm{MeV}$ and $\Gamma$ is the photon index.
	  	Integral flux above 100~MeV is given in the last column.
	  }  
\label{tab:spectra}
\end{table}


\section{Discussion}
\label{sec:discussion}

Particular attention was paid to the characteristics of the initial subject of investigation, the blazar \es, and the brightest of the new hotspots, S1.
We examine the spectral energy distributions of both sources in Section~\ref{sec:spectra}.
Changes of the their fluxes over time are inspected in Section~\ref{sec:lightcurves}.
Discussion of detection of S1 is given in Section~\ref{sec:detection}.
Finally, possible astrophysical counterparts to the unknown source candidates are reported in Section~\ref{sec:counterparts}.


\subsection{Spectral energy distributions}
\label{sec:spectra}

\begin{figure}
        \centering
        \begin{subfigure}[b]{0.49\columnwidth}
                \includegraphics[width=\textwidth]{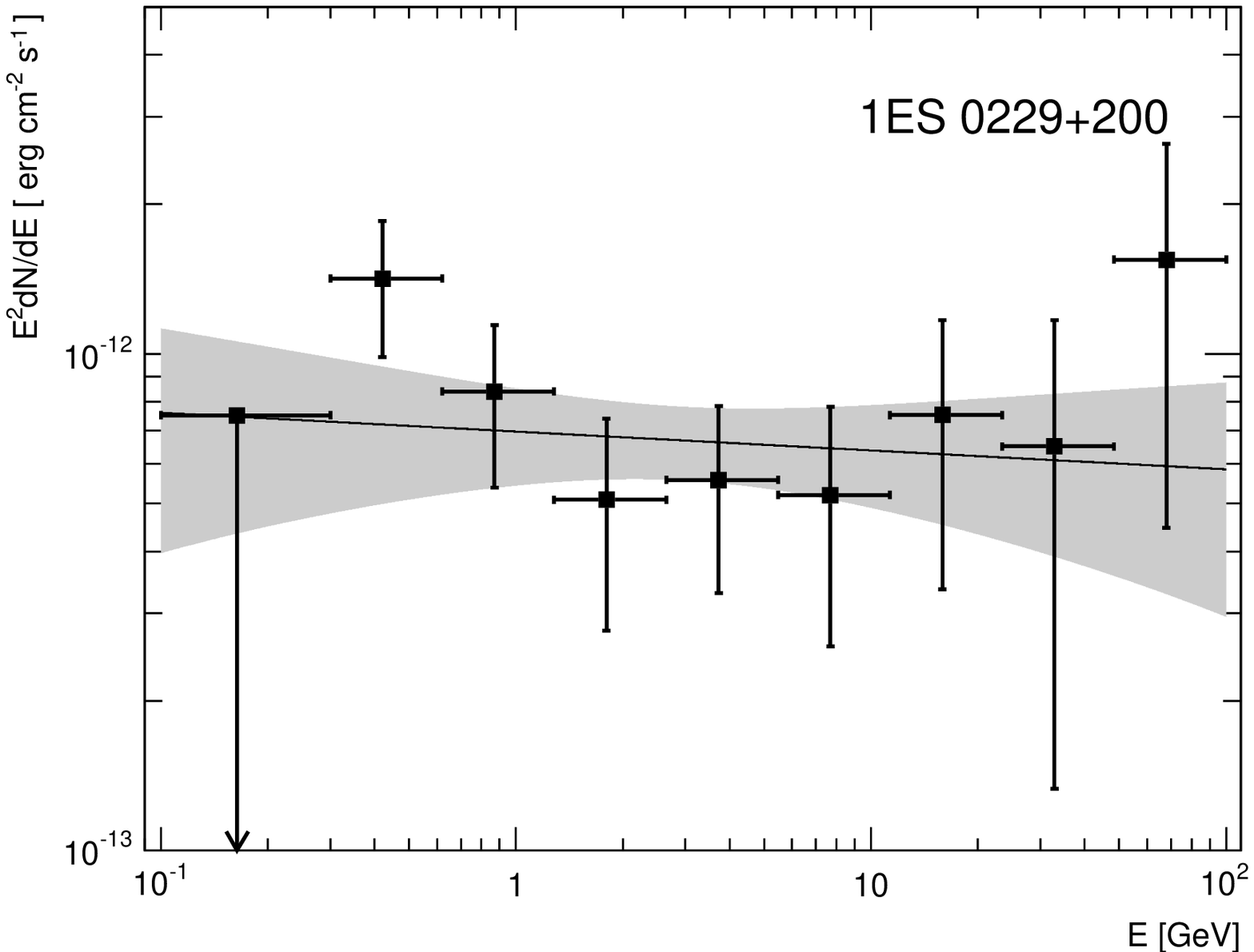}
                \subcaption{
                		\es
					}
                \label{fig:spectrum_1ES}
        \end{subfigure}
        \begin{subfigure}[b]{0.49\columnwidth}
                \includegraphics[width=\textwidth]{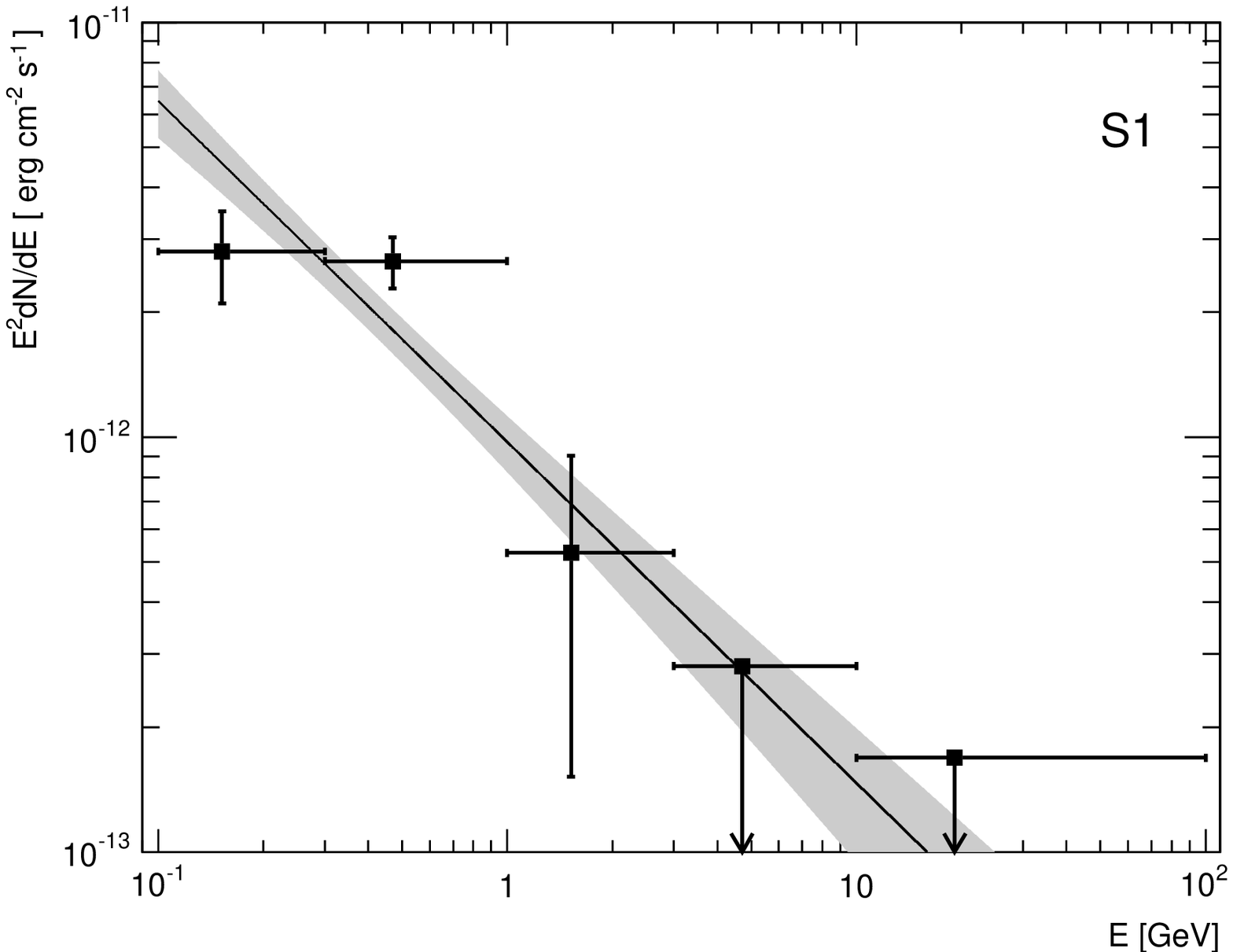}
                \subcaption{
                		S1
				}
                \label{fig:spectrum_unk}
        \end{subfigure}
        \caption{
        		Spectral energy distributions of \es~and S1.
			Upper limits on the flux at a $95\%$ confidence level denoted by arrows are stated in those bins where $\mathrm{TS}<1$.			
			Solid lines represent the spectral shapes given by maximum likelihood estimates of the relevant parameters obtained for the data from the whole studied energy range.
			Grey areas are defined by uncertainties on the differential fluxes.
			The error bars denote the statistical uncertainties at a $1\sigma$ confidence level.
		}
		\label{fig:sed}
\end{figure}

Spectral parameters of \es~deduced from over 6 years of data (see Table~\ref{tab:spectra}) are in good agreement with those reported by the \fermi~collaboration after 4 years of observations~\cite{3FGL}.
Spectral energy distribution of \es~is depicted in Fig.~\ref{fig:spectrum_1ES}.
The source is not detected below 300~MeV.
Maximum likelihood estimates of the spectral parameters obtained for the data within the range $0.1-100~\mathrm{GeV}$ suggest that the source exhibits a hard energy spectrum not inconsistent with the power law shape.
On closer inspection, we notice that the source spectrum is steeper in the range $0.3-2~\mathrm{GeV}$ than at higher energies.
The high energy part of the spectrum is compatible with the results of Vovk et al.~(2012) who studied the source in the range $1-300~\mathrm{GeV}$~\cite{vovk}.

The self--synchrotron Compton (SSC) scenario~\cite{katarzynski} is commonly employed to explain the origin of $\gamma$--radiation in blazars.
The one--zone SSC model~\cite{katarzynski} was used by the VERITAS collaboration to describe the available multiwavelength data~\cite{veritas}, including the results of previous analysis of \fermi~data~\cite{vovk}.
The GeV--TeV data were well described by a single peak attributed to the inversely Compton scattered population of photons (see Fig.~6 in Ref.~\cite{veritas}).
However, our results suggest that a break may be present in the spectrum around the energy of 2~GeV (see~Fig.~\ref{fig:spectrum_1ES}).
This may be a sign of a secondary component in the source $\gamma$--ray emission.

S1 is most active at energies below roughly 3~GeV, see Fig.~\ref{fig:spectrum_unk}.
The spectral analysis revealed that the test statistic corresponding to the detection of S1 above 3~GeV is $\mathrm{TS}<1$.
The signal is most significant in the energy range where the containment angle of photon directions reconstructed by \fermi~is the largest.
Thus, it is expectable that the extent of S1 is the largest among all new hotspots.
In Fig.~\ref{fig:tsmap2}, the lack of any residual signal in the surroundings of S1 (compare with Fig.~\ref{fig:tsmap1}) demonstrates that the addition of a hypothetical point--like object at the position of S1 can indeed fully account for the unknown signal.


\subsection{Temporal evolution}
\label{sec:lightcurves}

\begin{figure}
        \centering
        \begin{subfigure}[b]{0.49\columnwidth}
                \includegraphics[width=\textwidth]{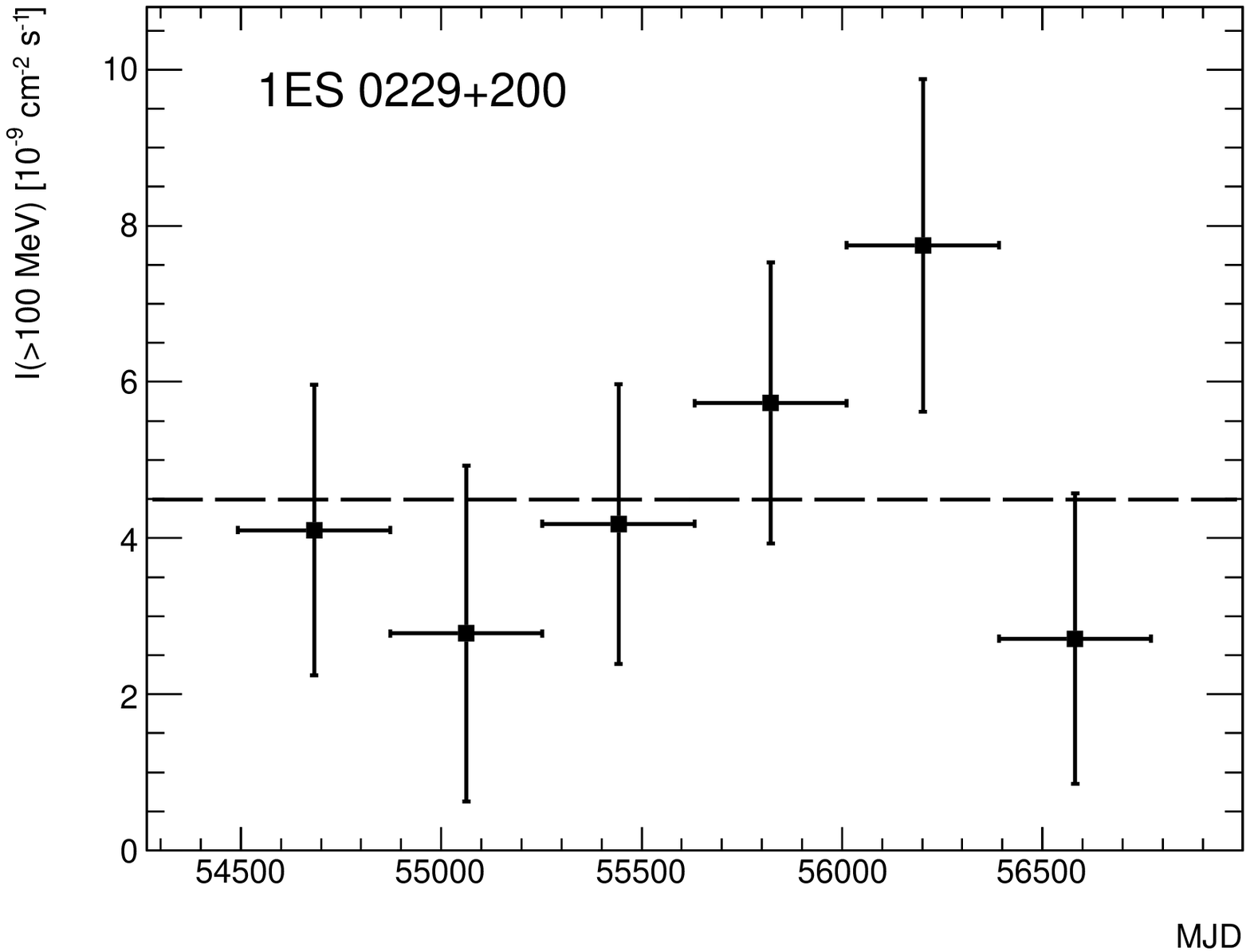}
                \subcaption{
                		\es
					}
                \label{fig:lightcurve_1ES}
        \end{subfigure}
        \begin{subfigure}[b]{0.49\columnwidth}
                \includegraphics[width=\textwidth]{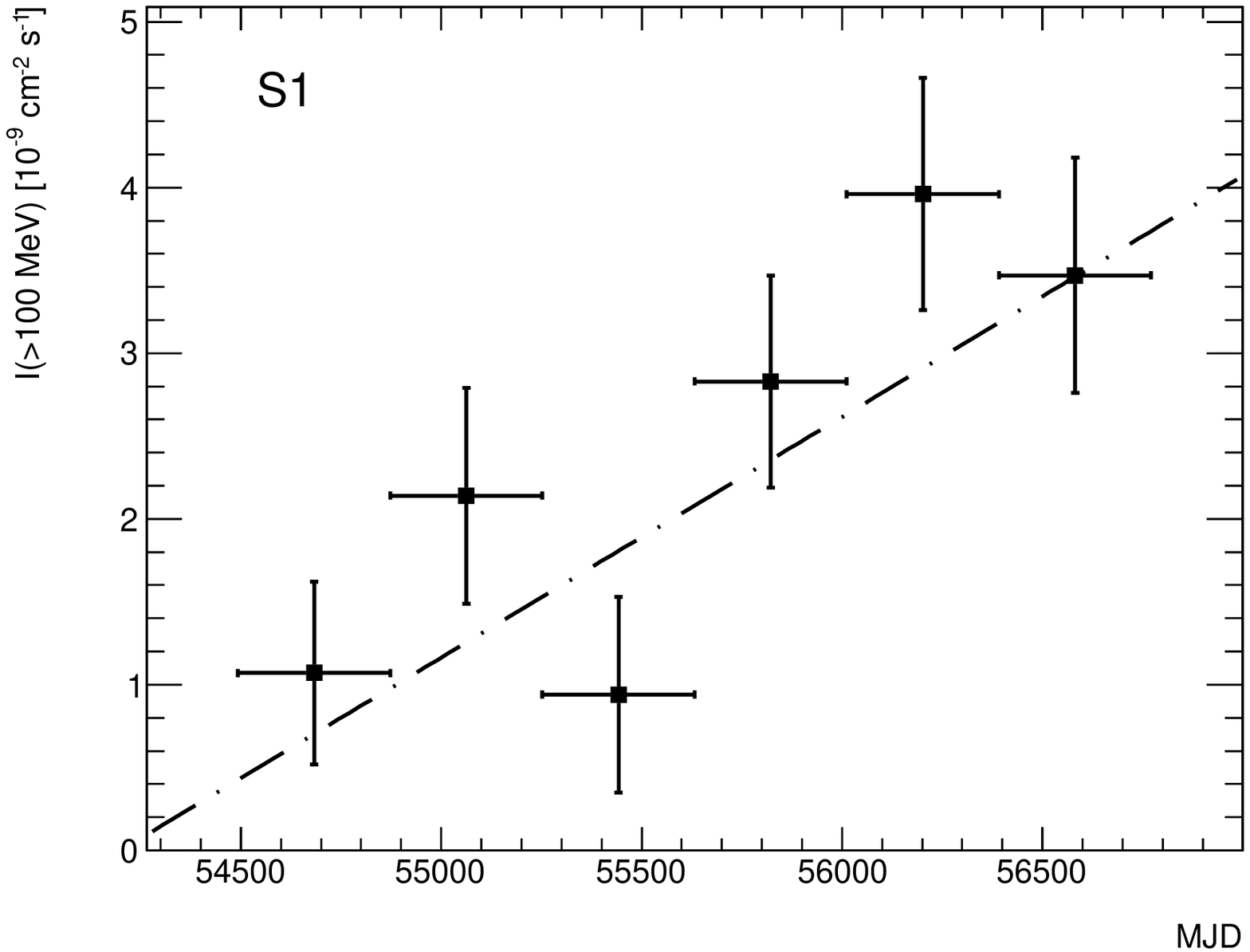}
                \subcaption{
                		S1
				}
                \label{fig:lightcurve_unk}
        \end{subfigure}
        \caption{
        		Lightcurves of \es~and S1.
			The dashed line represents the $\chi^{2}$ fit of the data from the direction of \es~to a constant.
			The dash--dotted line depicts the linear fit of S1 data.
			Statistical uncertainties at a $1\sigma$ confidence level are stated.
		}
		\label{fig:lightcurves}
\end{figure}

Temporal evolution of the integral fluxes of \es~and S1 is depicted in Fig.~\ref{fig:lightcurves}.
The whole investigated time range was divided into six equally long periods.
For each of them, integral fluxes were calculated assuming constant spectral indices $\Gamma$ as given in Table~\ref{tab:spectra}.

The $\chi^{2}$ fit of \es~data to a constant yields $\chi^{2} = 4.4$ for 5 degrees of freedom.
The probability that such or greater value of $\chi^{2}$ statistic is observed under the null hypothesis of stable flux is $P(\chi^{2},5) \approx 0.5$.
No significant deviations from the constant flux on a time scale of years can be recognized in the case of \es.

We note, however, that there has been some evidence for annual variability of this source in the very high energy (VHE) regime above 300~GeV~\cite{veritas,stefanik}.
In particular, the VHE intensity of the source was found to be in a higher state in the 2009--2010 season (MJD 55118--55212)~\cite{stefanik}.

The $\gamma$--ray intensity of S1 was changing over time, see Fig.~\ref{fig:lightcurve_unk}.
The $\chi^{2}$ fit of data from the direction of S1 to a constant (dashed line) yields a $\chi^{2}$ of 19.3 for 5 degrees of freedom corresponding to the $p$--value of $1.7 \times 10^{-3}$.
As the flux seems to follow an increasing trend over time, we fitted a linear function to the data points, see the dash--dotted line in Fig.~\ref{fig:lightcurve_unk}.
For the linear fit, we obtained a $\chi^2$ value of 6.1 for 4 degrees of freedom corresponding to the $p$--value of $0.2$.


\subsection{Detection of S1}
\label{sec:detection}

It is intriguing that though S1 comes out to be brighter than \es~in our analysis, S1 was not added to the 3FGL catalogue~\cite{3FGL} like the latter source.
We inspected this by analysing 4 years of data, i.e.~the time range examined by the \fermi~collaboration for the creation of the catalogue.

We used a source model containing all emitters from the 3FGL catalogue~\cite{3FGL} to test the alternative hypotheses that new sources \es~and S1 contribute to the signal in the studied region.
We obtained test statistics $\mathrm{TS}=48$ and $20$ for \es~and S1, respectively.
Thus, when taking into account contributions from all new source candidates in ROI, the test statistic for S1 is less than the minimal value of 25 that was set by the \fermi~collaboration for keeping the source candidates in the 3FGL catalogue~\cite{3FGL}.
It is also worth noting that the activity of S1 increased in the following years and the object became easier to recognize, see Fig.~\ref{fig:lightcurve_unk} and Table~\ref{tab:positions}.


\subsection{Counterparts}
\label{sec:counterparts}

We investigated the possible physical origin of $\gamma$--ray sources revealed in our analysis.
For that purpose, we looked for astrophysical objects which may be consistent with the positions of the suspected sources.
We found that two of the hotspots, S3 and S5, are very well compatible with the positions of BL Lac objects RXS~J02165+2314 and TEX~0239+175~\cite{VCV}, respectively.
These active galaxies are marked by green crosses in Fig.~\ref{fig:tsmap2}.
To our knowledge, neither of them has been reported to be active in the high energy regime so far.


\section{Conclusions}

We analysed 6.2 years of high energy $\gamma$--ray data gathered by the \fermi~experiment from a region surrounding the BL Lac \es.
The features of the source were found to be consistent with older findings~\cite{3FGL,vovk}.
We noticed a possible break in the spectrum around the energy of 2~GeV.
No increase of the source activity in the periods of the VHE enhancement was observed.

We found an unexpected significant signal near \es.
Additional four fainter unknown sources were recognized in the studied region.
None of the revealed objects have been previously reported.
Spectral shapes and fluxes of new sources of a $\gamma$--ray excess were recorded.
Possible correlation of two signal areas with BL Lacs RXS~J02165+2314 and TEX~0239+175 was found.
The brightest of the unknown signals was studied in detail.
No support for the extended nature of this candidate source has been established.
Its emission is variable and can be described by a steep power law.


\acknowledgments
This work was supported by the Czech Science Foundation grant 14--17501S.




\begin{thebibliography}{99}

\bibitem{3FGL}
Fermi--LAT Coll.,
\emph{Fermi Large Area Telescope Third Source Catalog},
(2015)
[\href{http://arxiv.org/abs/1501.02003}{\tt 1501.02003}]

\bibitem{PSF}
Fermi--LAT Coll.,
\emph{The Fermi Large Area Telescope On Orbit: Event Classification, Instrument Response Functions, and Calibration},
\emph{ApJS} {\bf 203} (2012) 4
[\href{http://arxiv.org/abs/1206.1896}{\tt 1206.1896}]

\bibitem{2FGL}
Fermi--LAT Coll.,
\emph{Fermi Large Area Telescope Second Source catalog},
\emph{ApJS} {\bf 199} (2012) 31
[\href{http://arxiv.org/abs/1108.1435}{\tt 1108.1435}]

\bibitem{vovk}
I. Vovk et al.,
\emph{Fermi/LAT Observations of 1ES~0229+200: Implications for Extragalactic Magnetic Fields and Background Light},
\emph{ApJL} {\bf 747} (2012) L14
[\href{http://arxiv.org/abs/1112.2534}{\tt 1112.2534}]

\bibitem{katarzynski}
K. Katarzy\'{n}ski et al.,
\emph{The multifrequency emission of Mrk 501. From radio to TeV gamma--rays},
\emph{A\&A} {\bf 367} (2001) 809--825

\bibitem{veritas}
E. Aliu et al.,
\emph{A Three-year Multi-wavelength Study of the Very-high-energy gamma-Ray Blazar 1ES~0229+200},
\emph{ApJ} {\bf 782} (2014) 13
[\href{http://arxiv.org/abs/1312.6592}{\tt 1312.6592}]

\bibitem{stefanik}
S. Stefanik, D. Nosek,
\emph{Variability of VHE gamma-ray sources},
\emph{Nucl.~Phys.~B--Proc.~Sup.} {\bf 256} (2014) 258--263
[\href{http://arxiv.org/abs/1412.2050}{\tt 1412.2050}]

\bibitem{VCV}
M.P. Veron--Cetty, P. Veron,
\emph{A catalog of quasars and active nuclei: 13th edition},
\emph{A\&A} {\bf 518} (2010) A10

\end{thebibliography}
\end{document}